\documentclass[a4paper,12pt]{article}

\usepackage{graphics}
\usepackage{latexsym}

\newcommand{\lsim}[1]{
\setlength{\unitlength}{12pt}
\begin{picture}(1.4,1.)
\put(.7,-0.3){\makebox(0.0,1.)[t]{$<$}}
\put(.7,-0.3){\makebox(0.0,1.)[b]{$\sim$}}
\end{picture}#1}

\begin{document}

\vspace{1cm}
\thispagestyle{empty}
\title{
\vspace{-2cm}
\begin{flushright}
{\normalsize IRB-TH-12/01 \\ 
\vspace{-0.5cm}
October 2001}
\end{flushright}
\vspace{1 cm}
\bf Renormalization-group running of the cosmological constant and its
implication for the Higgs boson mass in the Standard Model}

\vspace{1cm}
\author{A. Babi\'c\thanks{E-mail: ababic@thphys.irb.hr}, B. Guberina\thanks{E-mail: guberina@thphys.irb.hr},
 R. Horvat\thanks{E-mail: horvat@lei3.irb.hr},
 H. \v Stefan\v ci\'c\thanks{E-mail: shrvoje@thphys.irb.hr}
}

\vspace{2cm}
\date{
\centering
Rudjer Bo\v{s}kovi\'{c} Institute, \\
   P.O.Box 180, HR-10002 Zagreb, Croatia}


\maketitle

{\abstract 
The renormalization-group equation for the zero-point energies associated with
vacuum fluctuations of massive fields from the Standard Model is examined.
Our main observation is that at {\sl any} scale the running is necessarily
dominated by the heaviest degrees of freedom, in clear contradistinction
with the Appelquist \& Carazzone decoupling theorem. Such an enhanced running
would represent a disaster for cosmology, unless a fine-tuned relation
among the masses of heavy particles is imposed. In this way, we obtain 
$m_H \simeq 550 \;\mbox{\rm GeV}$ for the Higgs mass, a value safely within the 
unitarity bound, but far above the more stringent triviality bound  for the 
case when the validity of the Standard Model is pushed up to the grand 
unification (or Planck) scale. }

\vspace{1cm}

\vspace{1cm}

There are now increasing indications, based on observations on rich clusters
of galaxies \cite{1}, searches for Type Ia Supernovae \cite{2} and
measurements of the cosmic microwave background anisotropy \cite{3}, that
the today's universe is undergoing a phase of accelerated expansion. This is
usually attributed to the presence of a cosmological constant. Although the
simplest explanation is a time-independent (i.e. ``true'') cosmological
constant $\Lambda $, many scenarios have also been discussed involving a
dynamical cosmological constant $\Lambda (t)$. There have recently been a
number of suggestions regarding the nature of the latter, the most popular
candidate being known under the name of ``quintessence'' \cite{4} (a 
classically
unstable field that is rolling towards its true minimum which is presumed
to vanish).

The problem of the cosmological constant - how to reconcile its value from
cosmological observations, $\Lambda \sim \mbox{\rm 10}^{-47}\;\mbox{\rm
GeV}^4 $ (to be of the same order as the critical energy density) with
particle physics scales describing all known and unknown phase
transitions in the early universe or with $\Lambda_{Pl} \sim \mbox{\rm
10}^{72}\;\mbox{\rm GeV}^4 $ in the case of vacuum fluctuations with the
Planck scale cutoff - arises when an ordinary field theory is discussed in
relation to gravity. It is therefore  adequate to formulate the theory 
on the classical curved background \cite{5, Birrel}. However, it is true that the net
cosmological constant, being the sum of a certain number of essentially
disparate contributions, may classically always be set to zero by applying
some fine-tuning. It is to our current understanding that the problem is
intimately related to quantum gravity, leaving thereby string theory as the
only framework for properly addressing it \cite{6}.

In the two recent papers \cite{5, 7} Shapiro \& Sola found that even by
taking the quantum effects of the Standard Model, one could not fix the value
of the cosmological constant to any definite constant (including zero),
because any such a constraint would necessarily be invalidated at a
different scale (the energy scale changes in the course of the universe's
evolution) owing to renormalization group (RG) running effects. If the nature
of the RG behavior were such that near the scale $\mu = 0 $ one is allowed to
set $\Lambda (\mu = 0) = 0 $ (a relation suggesting some unknown symmetry
principle), then the above scenario could mimic quintessence models, but
without invoking a rolling scalar field. On the other hand, if one could not
set $\Lambda (\mu = 0) = 0 $, then the usage of the anthropic
principle would probably be the only alternative. It was argued in \cite{5,7} that the
scaling dependence of the cosmological constant should be consistent with
the standard cosmological model. That means that, given a value for $\Lambda
$ at far infrared, the running should reproduce the value for the
cosmological constant inferred from present observations at the present-day
scale ($\Omega_{\Lambda}^0 \simeq 0.6 - 0.7, \mu_0 \simeq 2 \times \mbox{\rm
10}^{-3}\;\mbox{\rm eV}$), and, also, should not spoil the success of
nucleosynthesis $(\Lambda \lsim \rho_R )$ at the much higher scale $\mu \sim
m_e $. Although the aim of Shapiro \& Sola in \cite{5, 7} was not to explain
fine-tuning to 55 decimals required to explain the present value of $\Lambda
$, careful examination of the running of $\Lambda $ could prove useful as  
it could reveal a close
relation between the SM parameters (particle masses and couplings) and the
parameters of observational cosmology.

The main result of \cite{5, 7} contains two nice features: (i) the running
near the present-day scale involves only light neutrino masses, and by
taking them to correspond to the large-mixing-angle $\rm MSW$ solution of the 
solar neutrino problem, we immediately arrive at the right value 
$\mid \Lambda \mid \approx \mbox{\rm 10}^{-47}\;\mbox{\rm GeV}^4$; (ii)
although the net value of the cosmological constant requires fine-tuning to
55 decimals, the running of the same quantity requires no fine-tuning at
all, thereby making its scaling dependence trivially consistent with the
standard cosmological scenario. There is, however, a bad feature too: in order
to set the condition $\Lambda (\mu = 0) = 0 $ (as to avoid the anthropic
principle for explanation of the observed values of cosmological parameters),
one should inevitably accept the existence of some light scalar with a mass a
few times the neutrino mass, which apparently 
leads us beyond the Standard Model. All
the above features stem from the fact that the authors of \cite{5, 7}
explicitly assumed the validity of the Appelquist \& Carazzone decoupling
theorem \cite{8}. In particular, this means that the quantum effects of some
particle are taken into consideration only at scales higher than the mass
of this particle $(\mu > m)$. As a consequence of this decoupling of heavy
particles, only
light neutrinos contribute to the running at present scales $\sim \mbox{\rm 10}^{-3}\;\mbox{\rm eV}$. 
In the present paper, we
scrutinize the decoupling theorem and its role in the running of 
$\Lambda $ in the Standard Model, and find that although the
contribution of a particle having a mass $m$ is suppressed at $\mu < m $,
it is still much larger than the contributions from lighter particles with
$\mu > m_i $. Thus, the heaviest particles do dominate the running at any
scale, and in order to have the RG behavior in accordance with standard
cosmology a fine-tuned relation connecting the heaviest masses should exist.
In this way, we obtain an interesting prediction for the Higgs mass in terms
of other particle masses in the Standard Model. 
Although the amount of fine-tuning in this relation is significantly
reduced in comparison with the original problem (55 decimals), it is still
considerable (28 decimals).

Let us start with the discussion of the cosmological constant $\Lambda$ 
which enters the Einstein equation in the following way:

\begin{equation}
R_{\mu\nu} - \frac{1}{2} g_{\mu\nu}R + 8\pi G g_{\mu\nu} \Lambda =
-8\pi G T_{\mu\nu} \, ,
\end{equation}

where $\Lambda$ is a dimensionful parameter with the dimension
$(mass)^4$. The classical general relativity does not bring any specific
preference for the value of $\Lambda$. Its value is therefore arbitrary.

With the advent of particle physics and quantum field theory it became clear
that $\Lambda$ can be interpreted as the vacuum energy density. In fact, 
there are additional sources of the cosmological constant coming from
particle physics. Field condensates at the classical level, and
zero-point energies at the quantum level, are two well-known generators of the
vacuum energy. Therefore, we have at least three sources of the cosmological
constant, 
1. the original Einstein constant,
2. field condensate contributions at the classical level,
3. particle zero-point energies at the quantum level.

The formulation of the theory \cite{Buchbinder, 5, Birrel} is rather simple - one constructs a 
renormalizable gauge theory (the gauged Higgs lagrangian, for example) in
an external gravitational field. In fact, one starts with the usual matter 
action in flat space-time, and replaces the partial derivatives by the 
covariant ones, the Minkowski metric by the general one, and $d^4 x$ by 
$d^4x\sqrt{-g}$. The cosmological constant $\Lambda$ that enters the Einstein
lagrangian may be regarded as a bare parameter, and used to absorb eventual
divergences coming from the quantum fluctuation in the particle lagrangian.
In such a way, the divergences of particle field theory are absorbed into
the bare $\Lambda$, and are therefore reduced to the geometry. It turns out
that, for example, the vacuum action necessary to insure the renormalizability
of the gauged scalar lagrangian should contain the terms 
$R_{\mu\nu\rho\sigma}^2$, $R_{\mu\nu}^2$, $R^2$, and $\Box R$, with
the corresponding coefficients $a_i$ which are the bare parameters. In this way, all 
divergences in the vacuum action can be removed by the appropriate 
renormalization of the matter fields, their masses and couplings, bare 
parameters $a_i$, $G_{bare}$, $\Lambda_{bare}$, and the nonminimal parameter $\xi_{bare}$
which enters the action via a term of the form $\xi \phi^{\dagger}\phi R$.

Generally, the scalar field $\phi$ with the potential energy $V(\phi )$ has 
the following contribution to the action:

\begin{equation}
S = \int d^4x \sqrt {-g} [ \frac{1}{2} g^{\mu\nu} (\partial_{\mu}\phi
\partial_{\nu}\phi ) - V(\phi )] \, .
\end{equation}

If $\phi_{vac}$ is the value of the field $\phi (x)$ which minimizes the 
potential $V(\phi )$, then the lowest state has $T_{\mu\nu} = g_{\mu\nu}
V(\phi_{vac})$, which is the classical scalar field contribution to the 
vacuum energy. As an example let us take the Higgs scalar field with the
potential $V(\phi ) = -m^2 \phi^{\dagger}\phi +
\lambda (\phi^{\dagger}\phi )^2$. Then the Higgs condensate contribution (at 
the classical level) to the cosmological constant is

\begin{equation}
\label{eq:cond}
\Lambda^{cond} = -\frac{m^4}{4 \lambda} \, .
\end{equation}

We shall turn to the discussion of the above expression later.

The second source of the contributions to the cosmological constant are quantum 
fluctuations (zero-point energy) of the free fields. 
Each free quantum field (in the case of bosonic fields being basically a 
collection of an infinite number of harmonic oscillators) 
contributes an infinite amount of the vacuum energy to the cosmological 
constant.

In the following we calculate and discuss the running cosmological constant
$\Lambda (\mu )$ for the case of a scalar field. It will turn out that the
decoupling theorem, although perfectly valid for Green
functions in the field theory, fails in the case of the calculation of the cosmological 
constant.

Using the dimensional regularization in $d=4+2\epsilon $ dimensions and the
$\rm MS$ renormalization scheme, one gets for the quantum fluctuations ( 
zero-point energy) of the scalar field

\begin{equation}
\label{eq:bubble}
ZPE = 
\frac{M^4}{64 \pi^2} \left( \frac{1}{\epsilon} + \gamma - \ln 4\pi + \ln\frac{M^2}
{\mu^2} - \frac{3}{2} \right) \, .
\end{equation}

Defining the relation between the bare ($\Lambda_{bare}$) and renormalized
($\Lambda$) quantities as

\begin{equation}
\label{eq:ren}
\Lambda_{bare} = \mu^{d-4} (\Lambda + z_{\Lambda} M^4 ) \, ,
\end{equation}

one gets for the counterterm $z_{\Lambda}$ 

\begin{equation}
\label{eq:zlambda}
z_{\Lambda} = -\frac{1}{64 \pi^2}\frac{1}{\epsilon} \, .
\end{equation}

It is straightforward to write down the renormalization group equation for 
$\Lambda$, which reads

\begin{equation}
\label{eq:runM}
(4\pi )^2 \mu \frac{\partial }{\partial \mu }\Lambda(\mu) = 
 \frac{1}{2} M^4 \, .
\end{equation}

Once derived, eq. (\ref{eq:runM}) should be valid for any value of $\mu$.
However, the relation (\ref{eq:runM}) has been derived using  the $\rm MS$ renormalization
scheme, which is a mass-independent renormalization scheme. It is well known that such a 
scheme does not give the correct mass behavior of the $\beta$ functions.
Therefore, the expression (\ref{eq:runM}) gives the correct behavior of
$\beta_{\Lambda} = \mu \frac{\partial \Lambda}{\partial \mu}$ only for 
$\mu \gg M$. For $\mu \ll M$, we would expect the decoupling
of the heavy particle with mass $M$, i.e. $\beta_{\Lambda}$ is expected to be approximately
zero.


However, it would be premature to claim the validity of the decoupling theorem \cite{8},
because on purely dimensional grounds, one expects the corrections 
of the type $\mu^2 /M^2$ \cite{Manohar,Pich} to be insufficient to suppress the quartic power
of the mass M in (\ref{eq:runM}). To be more precise, let us assume that there
are two scalar particles, one with a heavy mass $M$, and the other with
a light mass $m$. Then, the RGE becomes

\begin{equation}
\label{eq:runMm}
(4\pi )^2 \mu \frac{\partial }{\partial \mu }\Lambda(\mu) =
\frac{1}{2} M^4 + \frac{1}{2} m^4 
\end{equation}

at the scale $\mu$, $\mu \gg M, m$. However, for $m\ll \mu \ll M$,
 one would expect the heavy scalar to decouple with the suppression
 factor $\mu^2 /M^2$ and
 eq. (\ref{eq:runMm}) would have the form

\begin{equation}
\label{eq:lightheavy}
(4\pi )^2 \mu \frac{\partial }{\partial \mu }\Lambda(\mu) =
\frac{1}{2}a \frac{\mu^2}{M^2} M^4 + \frac{1}{2} m^4 \, ,   
\end{equation}

where $a$ is the number of order $\mathcal{O}(1)$. Obviously, the suppression factor
$\mu^2 /M^2$ is not sufficient to suppress the contribution of the heavy scalar,
since

\begin{equation}
 \mu^2 M^2 \gg m^4 
\end{equation}

and the heavy scalar does not decouple. The reason for such a peculiar behavior 
of the cosmological constant is its high dimensionality 
$(mass)^4$.

The calculation of
zero-mode contributions for a given massive field can be related to the 
evaluation of the vacuum bubble diagrams (diagrams without external legs). 
The aforementioned calculation results
in a divergent quantity which must be properly regularized. We shall consider
``cutoff'' regularization procedure for a bosonic degree of freedom (e.g. a real
scalar field) which is more suitable for our purposes since it displays the structure 
of divergences more clearly. Other
regularization schemes (e.g. dimensional regularization) yield equivalent
results. The treatment of fermionic degrees of freedom is completely analogous 
to the treatment of bosonic degrees of freedom up to the opposite sign. 
The zero-point energy of a real scalar field is \cite{Kapusta}      

\begin{equation}
\label{eq:div}
ZPE = \frac{1}{(4\pi)^2} A_{0}^4 + \frac{1}{2}\frac{1}{(4\pi)^2} 
\left[ A^2 m^2 - \frac{m^4}{4} -  \frac{m^4}{2} \ln \frac{A^2 + m^2}{m^2} \right] \, ,
\end{equation}

plus additional terms which vanish as $A \rightarrow \infty$, $A$ being the four dimensional cutoff.
The term $A_{0}^4$ corresponds to the zero-point energy in the massless limit.
Since we are dealing with the divergent quantity, a consistent procedure of renormalization 
must be invoked. Various divergent contributions have to be renormalized by adding appropriate counterterms.
Quartically and quadratically divergent terms have to be subtracted completely (i.e. the choice of 
counterterms is unique) while in the 
case of a logarithmically divergent term, the most general counterterm includes some scale dependence.
In order to examine the effects of mass thresholds, it is necessary to apply a renormalization scheme 
in which the counterterms are scale and mass dependent \cite{Manohar,Pich}. This requirement clearly disqualifies 
the most widely applied renormalization schemes, such as the $\rm MS$ or $\rm \overline{MS}$ schemes.
There exists a version of the $\rm MS$ scheme \cite{Binetruy} which incorporates the effects of mass
thresholds (named by its authors as decoupling subtraction). This scheme keeps the contributions of massive 
particles at scales above the mass, while it excludes them completely at the scales below the mass
and therefore implies a ``sharp cutoff'' approximation. However, it is also based on the assumption of validity
of decoupling of the massive field at low scales.  
This last feature is yet to be tested in the case of the cosmological constant. 
The subtraction scheme, on the other hand, meets the aforementioned demand. The counterterm in this scheme 
is obtained by setting some exterior scale (like the momentum squared) in the divergent Green function
to a predetermined value (usually referred to as a renormalization point). From eq. (\ref{eq:div}) it is clear that
in our case there is no exterior scale (we treat the mass $m$ as a parameter), so even the subtraction scheme 
cannot be applied directly. One possible way out of this predicament is to use a very general form of the counterterm and 
then limit its form by imposing some reasonable conditions on the running of relevant quantities (contribution to
the zero-point energy part of the cosmological constant). This approach leads to the following relation between
unrenormalized and renormalized zero-point energy parts of the cosmological constant:      

\begin{equation}
\label{eq:counter}
\Lambda_{bare} = \Lambda - \frac{1}{(4\pi)^2} A_{0}^4 - \frac{1}{2}\frac{1}{(4\pi)^2}
A^2 m^2 + \frac{1}{2}\frac{1}{(4\pi)^2} m^4 \ln \frac{A^2 + m^2}{\mu^2 g(\frac{m}{\mu})} \, ,
\end{equation}

where the function $\mu^2 g(\frac{m}{\mu})$ represents the general scale and mass dependence of 
the counterterm. The results of renormalization in the subtraction scheme \cite{Manohar,Pich} strongly
suggest the form of the counterterm determined by the function 

\begin{equation}
\label{eq:natural}
\mu^2 g \left( \frac{m}{\mu} \right) = \mu^2 + m^2 \, .
\end{equation}

We consider this choice the most natural and consequently use it in the concrete calculations in 
the rest of the paper. Nevertheless, one can perform a more general analysis starting from the rather 
general form of the counterterm. By introducing $x = m/\mu$, the running of the vacuum part of the 
cosmological constant becomes

\begin{equation}
\label{eq:run}
\mu \frac{\partial \Lambda}{\partial \mu} = \frac{1}{(4\pi)^2} \frac{m^4}{2}
\left[ 1 -\frac{1}{2} \frac{x g'(x)}{g(x)} \right] \, ,
\end{equation}

where the prime denotes the derivative with respect to $x$. 
Let us start from the general form of the counterterm determined by
the function

\begin{equation}
\label{eq:general}
g(x) = \sum_{l=-n}^{m} c_{l} x^{l} \, .
\end{equation}

Valuable information can be gained by considering the following interesting limits
of the expression governing the running of eq. (\ref{eq:run}):

\begin{eqnarray}
\label{eq:lim}
\lim_{x\to\infty} \left[ 1 -\frac{1}{2} \frac{x g'(x)}{g(x)} \right] 
& = & 1 - \frac{m}{2} + \frac{1}{2} \frac{c_{m-1}}{c_{m}} \frac{1}{x} 
+ \mathcal{O} \left( \frac{1}{x^2} \right) \, , \nonumber \\
\lim_{x\to0} \left[ 1 -\frac{1}{2} \frac{x g'(x)}{g(x)} \right]
& = & 1 + \frac{n}{2} + \frac{1}{2} \frac{c_{-n+1}}{c_{-n}} x 
+ \mathcal{O}(x^2) \, .
\end{eqnarray}

The first limit covers the behavior for large $x$, i.e. at scales $\mu$ much smaller than the mass $m$.
At low scales one expects suppressed contributions of very massive fields. If we formulate this 
expectation as a requirement, serious constraints on the index $m$ can be obtained. For $m \ge 3$, the running is 
unsuppressed and negative. The negative running at low scales, together with the positive running at higher scales, 
implies a change of sign at some intermediate scale which is clearly an undesirable property. For $m=0,1$, the running is positive,
but unsuppressed. Only for $m=2$, we obtain the suppressed behavior as required. In the opposite limit of small $x$, i.e.
large scales $\mu$ compared with the mass $m$, we demand to recover the behavior displayed by the $\rm MS$ and $\rm \overline{MS}$
schemes. Namely, in this limit,
the effect of mass threshold can be completely neglected, which is exactly the property of the $\rm MS$ and $\rm \overline{MS}$ schemes.
Therefore, the condition $n=0$ follows directly. Taking into account the considerations given above, the most general form 
of the counterterm (\ref{eq:general}) is given by $\mu^2 g(m/\mu) = \mu^2 + c_{1} m \mu + c_{2} m^2$ (the coefficient in front of
$\mu^2$ can be absorbed by the redefinition of the $\mu$ scale). Since terms linear in the mass $m$ are nonspecific for relativistic 
calculations, it is evident that our choice (\ref{eq:natural})
fits nicely into the allowed form of the counterterm.

Now when the question of the renormalization scheme is settled, we can look at the running of the vacuum part of the cosmological constant
in some particle physics model with its own spectrum of massive bosonic and fermionic degrees of freedom (relevant in our case).
The common property of the running in all models is the nonexistence of decoupling at low scales. Namely, 
for the contribution of the real scalar
field to the running of the zero-point energy part of the cosmological constant we obtain

\begin{equation}
\label{eq:realhigh}
\mu \frac{\partial \Lambda}{\partial \mu} =  \frac{1}{(4\pi)^2} \frac{1}{2} m^4
\end{equation}

in the $\mu \gg m$ limit. In the opposite $\mu \ll m$ limit, the running
becomes

\begin{equation}
\label{eq:reallow}
\mu \frac{\partial \Lambda}{\partial \mu} =  \frac{1}{(4\pi)^2} \frac{1}{2} m^2 \mu^2
\end{equation}

as anticipated in the relation (\ref{eq:lightheavy}). One can clearly see that the suppression 
of very massive fields is present, but insufficient to insure their decoupling.
 
In the case of the Standard Model, the running acquires the form 

\begin{equation}
\label{eq:standardrun}
(4\pi)^2 \mu \frac{\partial \Lambda}{\partial \mu} =
-2 \sum_{i} N_{i} m_{i}^4 \frac{\mu^2}{\mu^2 + m_{i}^2}
+ 3 m_{W}^4 \frac{\mu^2}{\mu^2 + m_{W}^2}
+ \frac{3}{2} m_{Z}^4 \frac{\mu^2}{\mu^2 + m_{Z}^2}
+ \frac{1}{2} m_{H}^4 \frac{\mu^2}{\mu^2 + m_{H}^2} \, ,
\end{equation}

where the index $i$ denotes fermions, $N_{i}$ being $3$ for quarks and $1$ for leptons. 
Direct integration of (\ref{eq:standardrun}) gives

\begin{eqnarray}
\label{eq:standardint}
(4\pi)^2 (\Lambda(\mu) - \Lambda(0)) & = &
- \sum_{i} N_{i} m_{i}^4 \ln \frac{\mu^2 + m_{i}^2}{m_{i}^2}
+ \frac{3}{2} m_{W}^4 \ln \frac{\mu^2 + m_{W}^2}{M_{W}^2} \nonumber \\
& & + \frac{3}{4} m_{Z}^4 \ln \frac{\mu^2 + m_{Z}^2}{M_{Z}^2}
+ \frac{1}{4} m_{H}^4 \ln \frac{\mu^2 + m_{H}^2}{m_{H}^2} \, .
\end{eqnarray}

The expression given above indicates that the contribution of very 
massive fields is nonnegligible at all scales. 
As for neutrinos, recent experiments indicate that 
neutrinos do have nonzero masses. 
The question of these masses is still unsettled, but it is
general agreement that they are in the region below $\mathcal{O}(1 \, \rm eV)$. As these masses are
far below all the other masses in play, we shall put them all to zero as a starting
approximation. The investigation of possible subtle effects due to nonzero neutrino masses 
will be discussed elsewhere.

In this framework we can focus our attention to the effects of running at scales rather below
the mass of the electron, the lightest particle in our approach. 
Since all the masses are large compared with the scale $\mu$, it is convenient to expand the logarithms in the 
relation (\ref{eq:standardint}). This procedure yields

\begin{equation}
\label{eq:runlow}
\Lambda(\mu) - \Lambda(0) = \frac{1}{(4 \pi)^2} \frac{1}{4} \mu^2 \left[ m_{H}^2 + 3 m_{Z}^2 + 6 m_{W}^2
- 4 \sum_{i} N_{i} m_{i}^2 \right] +  \frac{1}{(4 \pi)^2} \mu^4 \left[ \frac{1}{2} \sum_{i} N_{i} - \frac{5}{4} \right]
+ \mathcal{O} (\frac{\mu^6}{m_{large}^2}) \, .
\end{equation}

The analysis of the relation given above tells us instantly that large masses in the $\mu^2$ term drive 
the numerical value of $\Lambda$ far out of the range consistent with observation, unless the expression in
the brackets of the same term vanishes. Therefore, to avoid inconsistency with observation, we obtain a stringent
condition on the Higgs mass, i.e. an explicit expression for $m_{H}$ in terms of masses of other particles in the 
Standard Model:

\begin{equation}
\label{eq:mH}
 m_{H}^2 = 4 \sum_{i} N_{i} m_{i}^2 - 3 m_{Z}^2 - 6 m_{W}^2 \, .
\end{equation}

Using the numerical values from \cite{PDG} we obtain $m_{H} \simeq 550 \, \rm GeV$.
It is clear that the relation (\ref{eq:mH}) implies the relation between the squares 
of masses ranging from $\sim 1 \,\rm MeV$ to $\sim 100 \, \rm GeV$ and, accordingly, 
introduces a certain fine-tuning of masses of the Standard Model.

If the Higgs mass is fixed by the requirement (\ref{eq:mH}), the running of
the zero-point energy part of the cosmological constant is given by the $\mu^4$ term of the 
expression (\ref{eq:runlow}). The running below the electron mass should not be too 
intensive in order to preserve the conditions for nucleosynthesis. 
Using the expression for the energy density of radiation during nucleosynthesis
$\rho_{R} = \frac{\pi^2}{30} g_{*} T^4$, as well as making a natural choice for the scale
$\mu = T$, one obtains the relation

\begin{equation}
\label{eq:nucl}
\frac{\Lambda(\mu) - \Lambda(0)}{\rho_{R}} = \frac{555}{32 \pi^4 g_{*}} \, .
\end{equation}

With the numerical value $g_{*} = 3.36$, the expression given above 
acquires the numerical value $0.053$, a value within the range that 
does not disturb \cite{9} nucleosynthesis (note that an even more stringent constraint
obtained very recently in \cite{10} is obeyed). It is interesting to notice that in
the radiation dominated universe the quantity on the left-hand side of (\ref{eq:nucl})
is constant for $\mu \ll 1 \, \rm MeV$. This phenomenon of ``scaling'' has already been 
met in the investigations of scalar field cosmologies with potentials having attractor
solutions and its appearance here represents a very 
interesting and potentially important similarity. 

Finally, the relation (\ref{eq:runlow}) together with the constraint (\ref{eq:mH})
enables us to calculate the value of $\Lambda$ at present scale of the evolution of the universe.
If we take the value  $\mu_{0} = 0.002 \, \rm eV$ suggested in \cite{5,7},
we obtain $\Lambda(\mu) - \Lambda(0) \approx 10^{-48} \, \rm GeV^4$, a value reasonably close to the 
observed value of dark energy density of order $ 10^{-47} \, \rm GeV^4$. From (\ref{eq:runlow}) the amount 
of fine-tuning at present is estimated to be 1 in $(100 \, \rm GeV)^2/\mu_{0}^2 \approx 10^{28}$. 

At this point a remark is in order. All our preceding results have been obtained using a 
specific form of the function determining the counterterm $\mu^2 g(\frac{m}{\mu}) = \mu^2 + m^2$.
However, if we use a more general form $\mu^2 g(\frac{m}{\mu}) = \mu^2 + c_{2} m^2$, where $c_{2}$
is naturally expected to be of order $1$, the fine-tuning expression for the Higgs mass
(\ref{eq:mH}) remains completely the same, while the $\mu^4$ term stays of the same 
order of magnitude and the conclusions of comparisons of our results with observations
remain unchanged.

Let us now discuss our value ($\simeq 550 \;\mbox{\rm GeV}$) in view of
experimental and theoretical constraints on the mass of the Standard Model
Higgs boson. The experimental lower limit is $114 \;\mbox{\rm GeV}$
\cite{LEP} at $95\%$ confidence level, a value somewhat  higher than the ``best
fit'' value obtained from electroweak precision data \cite{Chivukula} ($106
\;\mbox{\rm GeV}$). At the same time, $m_H < 220 \;\mbox{\rm GeV}$ at $95\%$
confidence level. Thus, these data suggest that the Higgs boson mass should not
be too much larger than the present limit from direct searches. On
theoretical grounds, a well-known upper limit on the Higgs mass comes from
the unitarity of the scattering matrix. Even the most restrictive bound
$(\sim 800 \;\mbox{\rm GeV})$ obtained from the scattering process, $Z_L
W_L \rightarrow Z_L W_L $, is considerably higher than our value. On the
other hand, the triviality of the Standard Model admits it only as an
effective theory, valid below some energy scale $\Lambda $. If the validity
of the Standard Model is pushed up to extremely high scales (grand
unification or Planck), the triviality bound is more stringent than the
unitarity bound, being $\lsim 200 \;\mbox{\rm GeV}$ for the quartic coupling
taken to reside in the perturbative domain, $1 \lsim \lambda \lsim 10 $.
Thus, even on purely theoretical grounds, one can see (upon including the
stability lower bound) that $m_H $ in the $100 - 200 \;\mbox{\rm GeV}$ range is
preferred.

Since we take the heaviest masses from the Standard Model, our model implicitly
assumes the validity of the Standard Model up to the highest scales, thereby
violating the triviality bound. One can therefore claim, using arguments
based solely on the running of the cosmological constant, the existence of
some intermediate energy scale at which we should expect nonstandard 
phenomena to take place. Models in which nonstandard physics compensates 
the effect of a heavy Higgs (at the same time fitting precision data as good 
as the Standard Model) can be found, for example, in a recent review 
\cite{Peskin}. The above conclusion may however not be definite as, for the 
full treatment, one also needs to include scaling effects from the vacuum 
energy induced by a scalar Higgs potential of the Standard Model,
which we consider next.

The Higgs field will contribute to $\Lambda$
via the vacuum condensate. The contribution is 
given by eq. (\ref{eq:cond}) and the renormalization group equation reads

\begin{equation}
\label{eq:runcond}
\frac{d}{dt} \Lambda^{cond}(t) = 
-\frac{m^2}{2 \lambda}\frac{dm^2}{dt} + \frac{m^4}{4 \lambda^2}\frac{d\lambda}{dt} \, , 
\end{equation}

where $t = \ln \frac{\mu}{\mu_{0}}$ and $m_{H}^2 = 2 m^2$.

Again, one 
would expect the Higgs contribution at the scale $\mu \ll m_H$ to diminish
owing to the decoupling. Unfortunately, it is not very difficult to convince
oneself that the suppression factors are of the form $\mu^2/m_H^2$, and cannot 
compete with the overall $m_H^4$ factor. 
To see this, one inspects (\ref{eq:runcond}). It is clear that one has to calculate 
$dm^2/dt$ and $d\lambda /dt$ in, for example, the momentum subtraction 
scheme. Let us check, for instance, the contribution of the self-energy diagram
(the Higgs loop with two external Higgs legs) which contributes to the running 
of $m^2(t)$. The renormalized contribution is 
proportional to the $m_H^2$ multiplied by the following integral:

\begin{equation}
\label{eq:j}
j(m_{H}, \mu ) = \int_{0}^{1} dx \frac{\mu^2 x(1-x)}{m_{H}^2 + \mu^2 x(1-x)} \, .
\end{equation}

For $\mu \gg m_{H}$, the integral gives

\begin{equation}
\label{eq:jhigh}
j(m_{H}, \mu ) = 1 + O(m_{H}^2/\mu^2)
\end{equation}

and in the limit $\mu \ll m_{H}$, one gets

\begin{equation}
\label{eq:jlow}
j(m_{H}, \mu ) = \frac{1}{6} \frac{\mu^2}{m_{H}^2} + O(\mu^4/m_{H}^4) \, .
\end{equation}

Since  $dm^2/dt$ in (\ref{eq:runcond}) multiplies $m^2$, the overall mass factor is $m_H^4$
for $\mu \gg m_{H}$, whereas the suppression in (\ref{eq:jlow}) is not enough
to suppress it for $\mu \ll m_{H}$. We have checked that the same is true
for the one-loop, four-point function (a fermion box with four external Higgs legs). 
which contributes
to the running of $\lambda (t)$, and, again, the decoupling theorem fails in 
the calculation of the running of $\Lambda^{cond} (t)$. The examples discussed 
above indicate that the same behavior as in the running of the zero-point energy 
part of $\Lambda(\mu)$, is expected for $\Lambda^{cond}$ as well.

In conclusion, we made a study to demonstrate a scaling dependence of the
cosmological constant by showing that its observational value is not
preserved at different energy scales. The running due to one-loop vacuum
bubble graphs associated with massive fields from the Standard Model, is
obtained in a closed analytic form. We have noted that the quantum theory of
gravity plus matter truncated at the one-loop level is an adequate description
because of the nonrenormalizability of gravity. Contrary to the
expectation from the Appelquist \& Carazzone decoupling theorem, we have found
that more massive fields do play a dominate role in the running at any
scale. We have also indicated that the same behavior should persist in the
running of the cosmological constant induced by spontaneous symmetry
breaking through the Higgs mechanism. Owing to heavy masses involved in the
running, the standard cosmological scenario may be found in jeopardy unless
some fine-tuning is applied. As a result, we get the prediction 
$m_H \simeq 550 \;\mbox{\rm GeV}$ for the Higgs
mass. Since this mass is not favored
by the present constraints, one may consider our
results obtained from the running of the cosmological constant as an
independent indication that the Standard Model cannot be the full theory at
all scales. 
Finally, the effects described in this paper are a feature of any quantum field theory comprising
massive fields. Consequently, one expects the same type of relations between masses of 
that theory (stemming from the consistency with observation at low scales) in any extension
(e.g. SUSY, GUTs) of the Standard Model.

{\bf Acknowledgements}

We are grateful to John Ellis and Nikolai Uraltsev for helpful discussions.
This work was supported by the Ministry of
Science and Technology of the Republic of Croatia under the
contracts No. 00980102 and 00980204.

\end{document}